\documentclass[aps,twocolumn,prl,amssymb,floatfix,showpacs]{revtex4}

\usepackage{graphicx}
\usepackage{epstopdf}

\begin{document}
\title{Ultra-fast spin avalanches in crystals of molecular magnets in terms of magnetic detonation}

\author {M. Modestov} 

\affiliation {Department of Physics, Ume{\aa} University, SE-901 87
Ume{\aa}, Sweden}

\author{V. Bychkov} 

\affiliation {Department of Physics, Ume{\aa} University, SE-901 87
Ume{\aa}, Sweden}

\author{M. Marklund}

\affiliation {Department of Physics, Ume{\aa} University, SE-901 87
Ume{\aa}, Sweden}

\begin{abstract}
Recent experiments (Decelle et al., Phys. Rev. Lett. \textbf{102}, 027203
(2009), Ref. \cite{Decelle-09}) discovered an ultra-fast regime of spin avalanches in crystals of magnetic
magnets, which was three orders of magnitude faster than the traditionally studied
magnetic deflagration. The new
regime has been hypothetically identified as magnetic detonation. Here we
demonstrate the possibility of magnetic detonation in the crystals, as a front consisting of a leading shock and a zone of Zeeman energy release. We study the dependence of the
 magnetic detonation parameters on the applied magnetic field. We find that
the magnetic detonation speed only slightly exceeds the sound speed  in
agreement with the experimental observations.
\end{abstract}

\maketitle

Molecular magnetism is a rapidly developing interdisciplinary research area
within material science \cite{Gatteschi-review-03, Barco-review-05}. One of the widely investigated materials in the
subject is $\rm{Mn}_{12} $-acetate, with a high spin number ($S = 10$) and strong magnetic anisotropy \cite{Gatteschi-review-03, Barco-review-05,Sessoli-Nature-93, Friedman-PRL-96,
Thomas-Nature-96}. At sufficiently low temperature all the spins of the molecules
become oriented along an applied external magnetic field, thus occupying the ground
state (e.g. $S_{z} = 10$); in this state the magnetization reaches its saturation value.
When the magnetic field direction is switched to the opposite one, the
former ground state becomes metastable with an increased potential energy
(the Zeeman energy) and a barrier separating it from the new ground state. Active research on the subject demonstrated that spin-relaxation from the metastable to the ground state  often happens  in the form of a narrow front spreading in a sample with velocity of a few meters per second \cite{Suzuki-05, Hernandez-PRL-05, McHugh-07,
Garanin-Chudnovsky-2007, Hernandez-08, Villuendas-08, Modestov-2011}.
Still, all these works focused on magnetic deflagration, i.e. a front of energy release
propagating due to thermal conduction at velocities much smaller than
the sound speed.

However, in contrast to other studies, recent experiments by Decelle et al., Ref.
\cite{Decelle-09}, discovered a new fast regime of the magnetic avalanches
in $\rm{Mn}_{12}$-acetate with a  front velocity estimated to be
 (2000-3000) m/s,  which exceeds the typical magnetic
deflagration speed by three orders of magnitude. Though a limited number of sensors led to rather large uncertainty in measuring the front
velocity,  the experiments still indicated clearly that it was comparable to
the sound speed in the crystals. Furthermore, Decelle et al.
\cite{Decelle-09} hypothetically interpreted the new regime as
\textit{magnetic detonation}.
Although this hypothesis looked reasonable, it still
required much theoretical work to be justified. In particular, detonations in
combustion problems demonstrate a propagation velocity
larger than the sound speed by order of magnitude and destructively high pressure \cite{LL-Fluidmechanics, Law-book}. Besides, even in
combustion science, the phenomenon of deflagration-to-detonation transition
has remained one of the least understood processes for more than seventy years,
despite its extreme importance \cite{Law-book, Dorofeev-2011}. It is only recently that
a quantitative theoretical understanding of this process has been achieved
\cite{Bychkov-et-al-2005, Akkerman-et-al-2006, Bychkov-et-al-2008,
Valiev-et-al-2010}.

In this Letter we demonstrate the possibility of magnetic
detonation, in the form of a front with  a leading shock and a zone of Zeeman
energy release. We study the dependence of the
 magnetic detonation parameters on the applied magnetic field. We find that
the magnetic detonation speed is only slightly greater than the sound speed, in
agreement with experimental observations.

In line with the experiments \cite{Decelle-09}, we consider magnetic avalanches in  $\rm{Mn}_{12}$-acetate with the Hamiltonian
\begin{equation}
\label{eq1}
\mathcal{H} = - \beta S_{z}^{2} - g\mu_{B} H_{z} S_{z},
\end{equation}
suggested in \cite{Garanin-Chudnovsky-2007}. Here $S_{z}$  is the spin projection, $\beta \approx 0.65K$  is the magnetic anisotropy constant,
$g \approx 1.94$  is the gyromagnetic factor, $\mu_{B}$ is the Bohr magneton, and $H_{z} $ is the
external magnetic field. The Hamiltonian (1) determines the Zeeman energy
release and the energy barrier of the spin transition (in temperature units), which  depend on the
magnetic field as
\begin{eqnarray}
\label{eq2}
 & E_{a}&= \beta S_{z}^{2} - g\mu_{B}H_{z}S_{z} + \frac{g^2}{4\beta}\mu_{B}^{2}H_{z}^{2},\\
\label{eq3}
& Q&= 2g\mu _{B} H_{z} S_{z},
\end{eqnarray}
respectively, with $S_{z} =10$. The energy barrier decreases with the field while the Zeeman energy
increases linearly. Next, we consider a stationary magnetic detonation in a crystal
of molecular magnets. Similar to the theory of shocks in solids
\cite{Zeldovich-Raizer}, we adopt the reference frame of the detonation
front and find the conservation laws of mass, momentum and energy according to
\begin{eqnarray}
\label{eq4}
\rho_{0}D &=& \rho u,\\
\label{eq5}
P_{0} + \rho_{0} D^{2} &=& P + \rho u^{2},\\
\label{eq6}
\varepsilon_{0} + \frac{P_{0}}{\rho_{0}} + \frac{1}{2}D^{2} + Q &=&
\varepsilon + \frac{P}{\rho} + \frac{1}{2}u^{2} + Q a,
\end{eqnarray}
where the label 0 designates the initial state, $D$ is the detonation speed, $u$ is velocity produced by the detonation, $\varepsilon$ is thermal energy per molecule, and $a$ is the fraction of molecules in the metastable state. Here we neglect the thermal conduction, since this is a comparatively slow process.   In the theory of shock waves one introduces the volume per unit mass $V \equiv 1 / \rho$ instead of density. The conservation laws Eqs. (\ref{eq4})--(\ref{eq6}) have to be complemented by an equation of state. Following Ref.\ \cite{Zeldovich-Raizer}, we represent the pressure and energy of condensed matter at low temperature as a combination of elastic and thermal components according to
\begin{equation}
\label{eq7}
P = \frac{c_{0}^{2}}{V_{0} n}\! \left[ \left( \frac{V_{0}}{V}
\right)^{n}\!- 1 \right] \!+ \frac{\Gamma }{V} \frac{Ak_{B} T^{\alpha+1}} {(\alpha+1)\Theta_{D}^{\alpha}},
\end{equation}
\begin{equation}
\label{eq8}
\varepsilon = \frac{c_{0}^{2}}{n} \!\left\{ {\frac{1}{n -
1}{\left[ {\left( \frac{V_{0}} {V} \right)^{n - 1} \!\!\!- 1} \right]}\! +
\!\frac{V}{V_{0}} - 1} \right\}\! + \frac{Ak_{B} T^{\alpha+1}} {(\alpha+1)\Theta_{D}^{\alpha}},
\end{equation}
where $c_0$ is the sound speed (we take $c_0\approx 2000\ $m/s  in accordance to \cite{Decelle-09}), the power exponent $n\approx 4$ as suggested in \cite{Zeldovich-Raizer}, $\Gamma \approx 2$ is the Gruneisen coefficient, $\Theta_{D}$ is the Debye temperature
with $\Theta_{D} = 38K$ for $\rm{Mn}_{12}$, $k_{B}$ is the Boltzmann constant, $A = 12\pi^{4} / 5$ corresponds to the simple crystal model,
$\alpha=3$ is the problem dimension.
Thus, in Eqs. (\ref{eq4})--(\ref{eq8}) we have a complete system for describing magnetic detonation in molecular magnets.

The properties of  shocks and detonations are represented by the Hugoniot/detonation curve $P=P(V)$, see Ref. \cite{LL-Fluidmechanics, Law-book, Zeldovich-Raizer}. We also introduce the scaled density ratio $r = \rho / \rho_{0} = V_{0}/V$ which characterizes the matter compression. Using Eqs. (\ref{eq4})-(\ref{eq8}), we derive the following implicit form for the detonation relation
\begin{eqnarray}
\label{eq9}
&& \left( \frac{1}{\Gamma}  - \frac{r - 1}{2}
\right)\frac{P}{\rho _{0}} =  \,rQ\left( {1 - a} \right) + \left( r + \frac{r - 1}{2}\Gamma \right)\varepsilon_{0} \nonumber\\
&& +\frac{c_{0}^{2}}{n - 1} \left[ {r - 1 -
\left( {1 - \frac{n-1}{\Gamma}}\right)\frac{r^{n} - 1}{n}} \right].
\end{eqnarray}
In the case of zero energy release ($a=1$), Eq. (\ref{eq9}) reduces to the Hugoniot equation  for a shock wave (which we denote by the subscript $s$). In the detonation, the leading shock compresses the sample, increases temperature and hence facilitates the spin reversal with the Zeeman energy release. The released Zeeman energy provides expansion of the medium, which acts like a piston and supports the leading shock. In the case of the completed spin reversal ($a=0$), Eq. (\ref{eq9}) describes the final state behind the detonation front (which we denote by the subscript $d$).
\begin{figure}
\includegraphics[width=3.4in,height=2.5in]{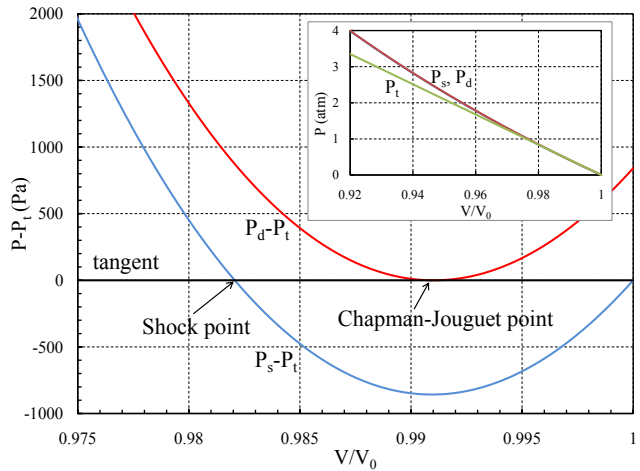}
\caption{The insert: Traditional presentation of the
Hugoniot and detonation curves and the tangent line to the detonation curve
in $\rm{Mn}_{12}$-acetate for the external magnetic field $H_{z}=4$T.
The main plot: The Hugoniot and detonation curves with the tangent line
extracted; label "t" stands for tangent.}
\end{figure}
The insert of Fig. 1 shows the Hugoniot and detonation curves found using Eq. (\ref{eq9}) for $H=4$ T. We assume that there is no external atmospheric pressure and the initial temperature is negligible, which corresponds to the initial point ($V=V_0;\, r=1;\ P=0$). Because of the energy release, the detonation curve is always above the Hugoniot one. In the case of $\rm{Mn}_{12}$ we find  that the elastic contribution to the pressure and energy dominates over the thermal one, which leads to a rather weak detonation with the shock and detonation curves almost coinciding as shown at the inset of Fig. 1. A self-supporting detonation corresponds to the Chapman-Jouguet (CJ) regime, for which velocity of the products in the reference frame of the front is equal to the local sound speed \cite{LL-Fluidmechanics}. The CJ point at the detonation curve is determined by the tangent line connecting the initial state and the detonation curve. Since the detonation and Hugoniot curves are extremely close at the insert of Fig. 1, the intersection of the tangent line cannot be seen in the traditional representation of the curves. In order to make the figure illustrative, we subtract this tangent line from the Hugoniot and detonation curves in Fig. 1. In the new representation, the tangent line corresponds to the zero line, while the Hugoniot and  detonation curves may be distinguished quite well. The CJ point in Fig. 1 corresponds to the final state behind the detonation front. The shock point indicates the strength of the leading shock as determined by the Zeeman energy release for the CJ regime; the density and the  pressure acquire maximum values at the shock front. The Zeeman energy release behind the shock produces expansion of matter with an ensuing pressure reduction.
	
 We notice from Fig. 1 that the $\textrm{Mn}_{12}$ crystal is compressed  by few percents in the detonation wave, which makes  an analytical theory for the detonation front parameters possible using expansion $r=1+\delta$ with $\delta\ll1$.  Then, to leading order in $\delta$, Eqs. (\ref{eq7}), (\ref{eq9}) reduce to
\begin{eqnarray}
\label{eq10}
P &=& \rho_{0} \left[ {\Gamma Q (1-a) + c_{0}^{2} \delta}  \right],\\
\label{eq11}
T^{\alpha + 1} &=& \left( {\alpha + 1} \right) \frac{\Theta_{D}^{\alpha
}}{Ak_{B}} \left[ {Q (1-a) +  \frac{n + 1}{12}c_{0}^{2} \delta ^{3}}\right].
\end{eqnarray}
We find the final compression behind the detonation front as
\begin{equation}
\label{eq12}
\delta_{d} = \frac{1}{c_{0}} \sqrt {\frac{2\Gamma Q}{n + 1}}.
\end{equation}
The compression behind the leading shock is larger by a factor of 2, $\delta_s \approx 2 \delta_d$, as may be seen from the parabolic shape of the Hugoniot and detonation curves in Fig. 1.
The detonation speed may be found from Eq. (\ref{eq4}) as
\begin{equation}
\label{eq13}
D = c_{0} \left( {1 + \frac{n + 1}{4}\delta_{s} }\right),
\end{equation}
which means that the magnetic detonation speed slightly exceeds the sound speed   in agreement with the experimental observations \cite{Decelle-09}. Substituting $\delta_s$, $\delta_d$, and $a=1; \, 0$ into Eqs. (\ref{eq10}), (\ref{eq11}), we find the analytical formulas for pressure and temperature at the shock and behind the detonation front, respectively. Taking into account Eq. (\ref{eq3}), these formulas specify the dependence of the detonation parameters on the external magnetic field.
In Fig. 2 we compare the analytical theory to the numerical solution to Eq. (\ref{eq9}). The curves of Fig. 2a show the density at the shock wave and behind the detonation front, when all the spins have been aligned along the magnetic field. We see that the density at the shock increases by less than  3 percents if the magnetic field is below 10 T. Because of this small compression, the analytical theory (\ref{eq12}) is in a very good agreement with the numerical solution. The maximum value of shock pressure is about 1.2 atm for 10 T. Thus, due to the small compression and the moderate pressure increase, the magnetic detonation does not destroy the magnetic properties of the crystals.
\begin{figure}
\includegraphics[width=3.4in,height=2.45in]{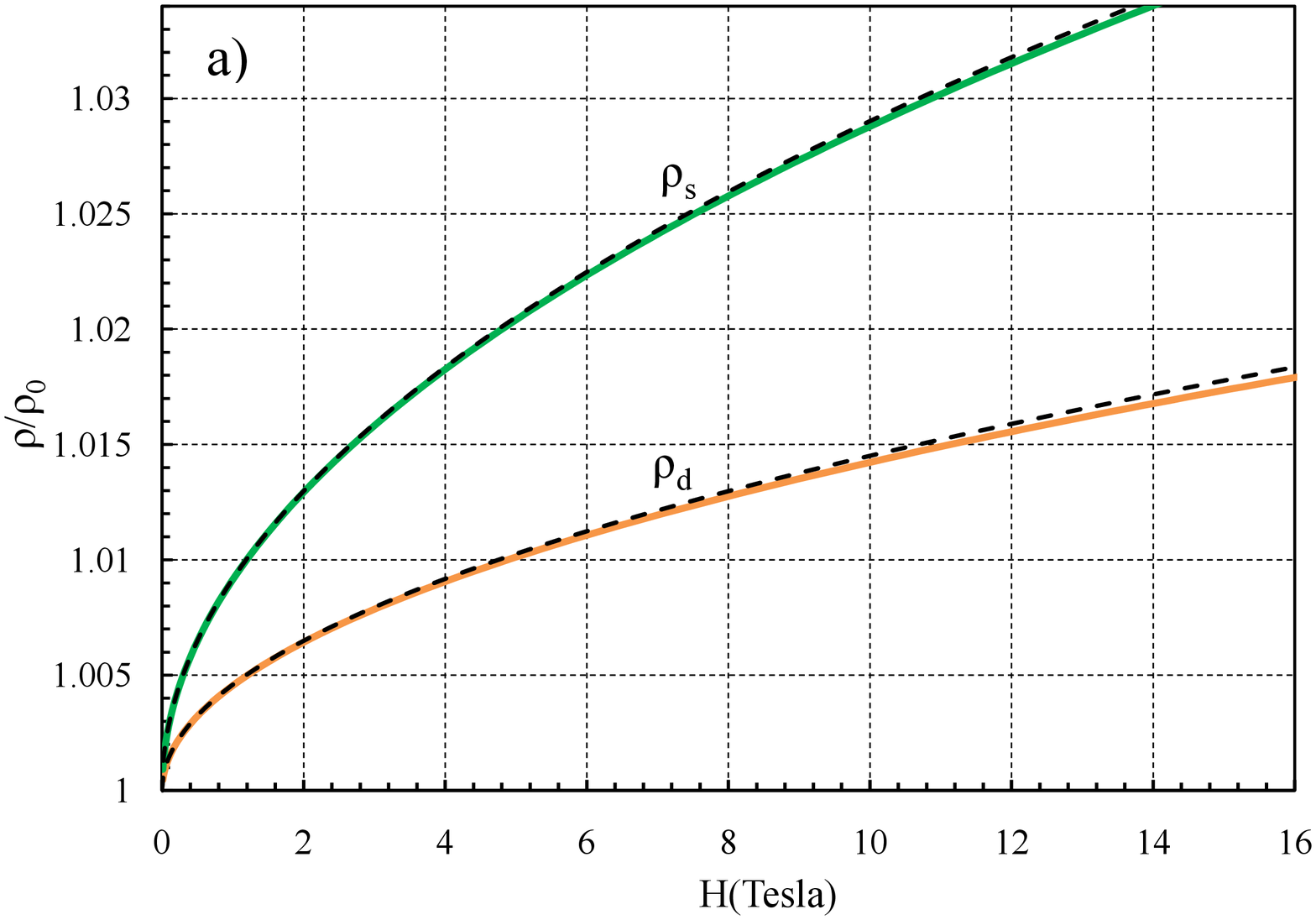}
\includegraphics[width=3.4in,height=2.45in]{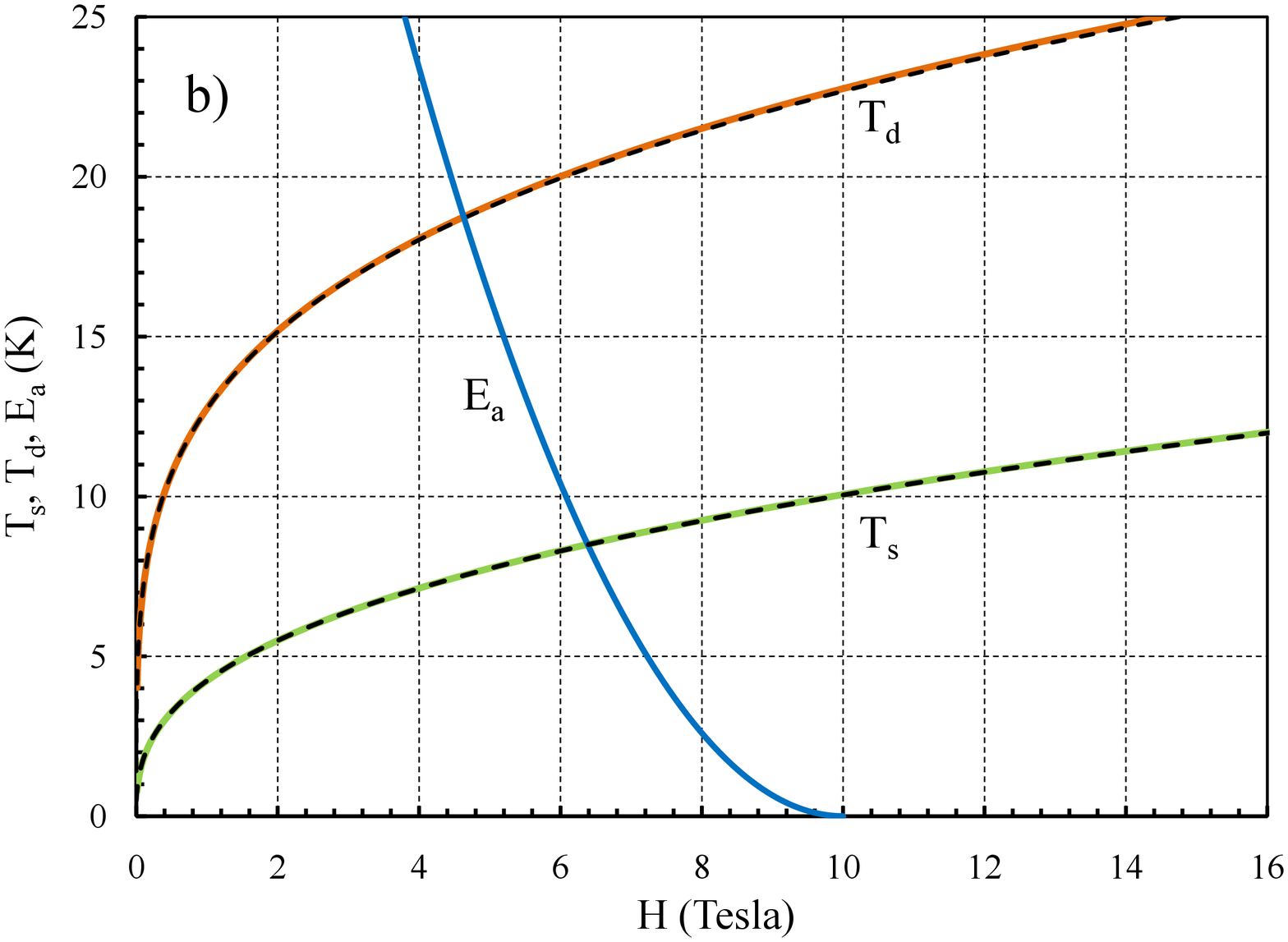}
\caption{Density ratio (a) and temperature (b) at the leading shock and behind the detonation front versus
the external magnetic field. Solid lines show exact numerical solution; the
dashed lines stand for the analytical theory.}
\end{figure}	

At the same time, the crystal temperature increases considerably because of the shock, and this stimulates a fast spin reversal and a further temperature increase. Fig. 2b illustrates the temperature increase at the shock wave and behind the detonation front. Again, we observe very good agreement between the analytical theory and the numerical solution. The temperature at the leading shock is comparable to that expected for the magnetic deflagration \cite{Suzuki-05, Hernandez-PRL-05,McHugh-07,Hernandez-08}, which also makes the reaction time  comparable in both processes. In  classical combustion, the temperature at the leading shock in the detonation wave is still quite small in comparison with the activation energy of the chemical reactions, so that the active reaction zone lags considerably behind the shock \cite{Law-book}. The situation may be quite different in  magnetic detonation. When the magnetic field is stronger than $2-3$ T, the shock temperature is relatively high ($E_a/T_s<5$) so that active spin reversal starts right at the shock wave. Figure 2b presents also the energy barrier as a function of the external magnetic field. The energy barrier decreases with the growth of the magnetic field, Eq. (\ref{eq2}), as shown in Fig. 2b. When the magnetic field exceeds 10 T, the energy barrier vanishes,  the metastable state turns unstable, and the molecules may settle down freely to the ground state. Hence, one may interpret magnetic avalanches as detonation or deflagration only for the fields below 10 T.
	
Finally, we describe the internal structure of the magnetic detonation front. In the reference frame of the moving front, the molecule fraction with the spin opposite to the field direction is determined by \cite{Garanin-Chudnovsky-2007}
\begin{equation}
\label{eq14}
u{\frac{\partial a}{\partial x}} = \frac{a}{\tau _{R}}\exp \left(
{ - \frac{E_{a}}{T}} \right),
\end{equation}
where $\tau_R$ is a constant of time dimension characterizing the spin reversal. We integrate Eq. (\ref{eq14}) numerically together with Eqs. (\ref{eq4}) and (\ref{eq7}) along the tangent line in Fig. 1, from the shock to the CJ point; the obtained profiles are depicted in Fig. 3 for $H=3$ T.
\begin{figure}
\includegraphics[width=0.95\columnwidth]{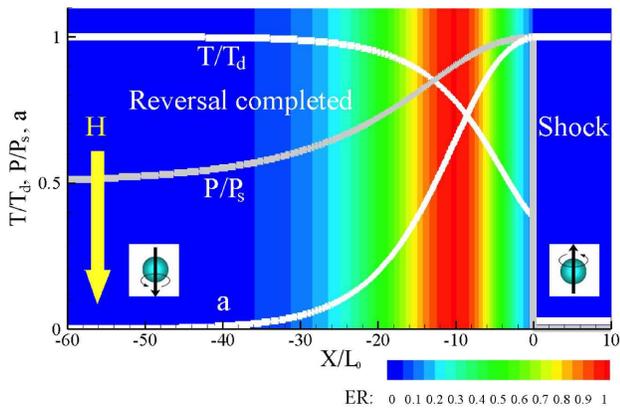}
\caption{Stationary profiles of the scaled temperature, pressure,
and fraction of molecules in the metastable state for $H_{z}=3$ T. The
background shading shows the energy release.}
\end{figure}	
The background shading represents the energy release due to the spin reversal; the temperature and the pressure are scaled to their maximal values. The coordinate is scaled by the characteristic length $L_0=c_0\tau_R \approx 2 \cdot 10^{-4} \textrm{m}$, where we take $\tau_R \approx 10^{-7}\textrm{s}$ as obtained in several experiments \cite{Suzuki-05, Hernandez-PRL-05, McHugh-07, Hernandez-08}. Using this value we can estimate the characteristic width of the stationary detonation wave to a few millimeters. The applied magnetic field influences strongly the reaction rate and thus the front width. For magnetic fields higher than 5 T, the detonation width is $< 1$ mm, while for a weaker field the width may increase considerably. For this reason, the detonation mechanism in molecular magnets may only be observed in  experiments utilizing  high enough magnetic fields, since the typical sample size is of order of several millimeters. The typical scales in the experiments of Ref. \cite{Decelle-09} were also about a few millimeters.  Thus the experimentally observed fast avalanche regime   was, presumably, a non-stationary detonation in the process of developing.

To summarize, in this Letter we have developed a theory of magnetic detonation in molecular magnets, which explains a new regime of ultra-fast spin avalanches discovered recently in the experiments of Ref. \cite{Decelle-09}. The detonation regime is two to three orders of magnitude faster than the magnetic deflagration observed before \cite{Suzuki-05, Hernandez-PRL-05, McHugh-07, Hernandez-08}. We have shown that the leading shock triggers the spin reversal in these magnetic  systems, and that the magnetic detonation propagates with velocities slightly larger than the sound speed. In contrast to traditional detonations in combustion, which are characterized by strongly supersonic velocities and ultra-high pressure, magnetic detonations involve rather moderate pressure increase, which is about 1 atm even for considerable magnetic fields. For this reason, magnetic detonation does not destroy magnetic properties of the crystals, a very important conclusion in view of possible applications of molecular magnets to, e.g., quantum computing.



This work was supported by the Swedish Research Council and by the Kempe
Foundation. The authors thank Petter Minnhagen, Bertil Sundqvist, Thomas
W{\aa}gberg, Sune Pettersson, Tatiana Makarova and Valeria Zagainova for useful discussions.


\begin{thebibliography}
\bibliographystyle{}

\bibitem{Decelle-09}
W. Decelle, J. Vanacken, V. V. Moshchalkov, J. Tejada, J. M. Hernandez, and
F. Macia, Phys. Rev. Lett. \textbf{102}, 027203 (2009).

\bibitem{Gatteschi-review-03}
D. Gatteschi and R. Sessoli, Angew. Chem., Int. Ed. \textbf{42}, 268 (2003).

\bibitem{Barco-review-05}
E. del Barco, A. D. Kent, S. Hill, J. M. North, N. S. Dalal, E. Rumberger,
D. N. Hendrikson, N. Chakov, and G Christou, J. Low Temp. Phys.
\textbf{140}, 119 (2005).


\bibitem{Sessoli-Nature-93}
R. Sessoli, D. Gatteschi, A. Caneschi, and M. A. Novak, Nature (London)
\textbf{365}, 141 (1993).

\bibitem {Friedman-PRL-96}
J. R. Friedman, M. P. Sarachik, J. Tejada, and R. Ziolo, Phys Rev. Lett.
\textbf{76}, 3830 (1996).

\bibitem{Thomas-Nature-96}
L. Thomas, F. Lionti, R. Ballou, D. Gatteschi, R. Sessoli, and B. Barbara,
Nature (London) \textbf{383}, 145 (1996).


\bibitem{Suzuki-05}
Y. Suzuki, M. P. Sarachik, E. M. Chudnovsky, S. McHugh, R. Gonzalez-Rubio,
N. Avraham, Y. Myasoedov, E. Zeldov, H. Shtrikman, N. E. Chakov, and G.
Christou, Phys. Rev. Lett. \textbf{95}, 147201 (2005).

\bibitem{Hernandez-PRL-05}
A. Hernandez-Minguez, J. M. Hernandez, F. Macia, A. Garcia-Santiago, J.
Tejada, and P. V. Santos, Phys. Rev. Lett. \textbf{95}, 217205 (2005).

\bibitem{McHugh-07}
S. McHugh, R. Jaafar, M. P. Sarachik, Y. Myasoedov, A. Finkler, H.
Shtrikman, E. Zeldov, R. Bagai, and G. Christou, Phys. Rev. B \textbf{76},
172410 (2007).

\bibitem{Hernandez-08}
A. Hernandez-Minguez, F. Macia, J. M. Hernandez, J. Tejada, and P. V.
Santos, J. Magn. Magn. Mater. \textbf{320}, 1457 (2008).

\bibitem{Garanin-Chudnovsky-2007}
D. A. Garanin and E. M. Chudnovsky, Phys. Rev. B. \textbf{76}, 054410
(2007).

\bibitem{Villuendas-08}
D. Villuendas, D. Gheorghe, A. Hernandez-Minguez, F. Macia, J. M.
Hernandez, J. Tejada, R. J. Wijngaarden, EPL (Europhysics Letters)
\textbf{84}, 67010 (2008).


\bibitem{Modestov-2011}
M. Modestov, V. Bychkov, and M. Marklund, accepted in Phys. Rev. B.

\bibitem{LL-Fluidmechanics}
L. Landau and E. Lifshitz, \textit{Fluid Mechanics}, Pergamon Press, Oxford, 1989.

\bibitem{Law-book}
C.K. Law, \textit{Combustion Physics}, Cambridge University Press, NY, 2006.

\bibitem{Dorofeev-2011}
S. Dorofeev, Proc. Combust. Inst. \textbf{33} 2161 (2011).

\bibitem{Bychkov-et-al-2005}
V. Bychkov, A. Petchenko, V. Akkerman, L.-E. Eriksson, Phys. Rev. E
\textbf{72}, 046307 (2005).

\bibitem{Akkerman-et-al-2006}
V. Akkerman, V. Bychkov, A. Petchenko, L.-E. Eriksson, Combust. Flame
\textbf{145}, 206 (2006).

\bibitem{Bychkov-et-al-2008}
V. Bychkov, D. Valiev, L.-E. Eriksson, Phys. Rev. Lett. \textbf{101}, 164501
(2008).

\bibitem{Valiev-et-al-2010}
D. Valiev, V. Bychkov, V. Akkerman, C. K. Law, L.-E. Eriksson, Combust.
Flame \textbf{157}, 1012 (2010)

\bibitem{Zeldovich-Raizer}
Ya. B. Zeldovich and Yu. P. Raizer, \textit{Physics of Shock Wave and High-Temperature Hydrodynamic Phenomena}, Dover Publications, Inc. Mineola, New
York (2002).

\end{thebibliography}
\end{document}